# Adaptive Modulation (QPSK, QAM)

*Rao Farhat Masood, Member IEEE, MIE (Pak), PEC*
*National University of Sciences and Technology, Pakistan.*

**Abstract**
*In this paper, introduced below are the concepts of digital modulation used in many communication systems today. Techniques described include quadrature phase shift keying (QPSK) and quadrature amplitude modulation (QAM) and how these techniques can be used to increase the capacity and speed of a wireless network. These modulation techniques are the basis of communications for systems like cable modems, DSL modems, CDMA, 3G, Wi-Fi\* (IEEE 802.11) and WiMAX\* (IEEE 802.16)*

**Carrier Waves**

Radio waves are electromagnetic waves that move at the speed of light in a sine wave formation and can be used to carry a message over a distance. They can have different frequencies which describes how fast they are moving up and down which is measured in cycles per second or hertz. Carrier waves with different frequencies have different properties. For example, light waves are visible to the naked eye but cannot travel through walls. Radio waves (especially those of lower frequency) can penetrate walls and buildings as well as bending (diffraction) around corners.

**Modulation**

Modulation is the process by which a carrier wave is able to carry the message or digital signal (series of ones and zeroes). There are three basic methods to this: amplitude, frequency and phase shift keying. Higher orders of modulation allow us to encode more bits per symbol or period (time).

**Amplitude Shift Keying (ASK)**

Amplitude shift keying (ASK) involves increasing the amplitude (power) of the wave in step with the digital signal (in other words, low = 0, high = 1) and is used in AM radio.

**Frequency Shift Keying (FSK)**

Frequency shift keying (FSK) changes the frequency in step with the digital signal. Systems that use this modulation (broadcast FM radio) tend to be more resilient to noise since noise usually changes the amplitude of the signal. In Figure 1, different bits are represented by different frequencies which can then be detected by a receiver.

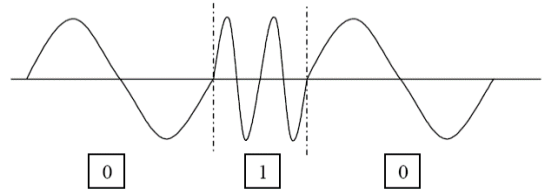

**Figure 1: Frequency Shift Keying (FSK)**

**Phase Shift Keying (FSK)**

Phase shift keying (PSK) changes the phase of the carrier in step with the digital message. For binary phase shift keying (BPSK), each symbol could indicate two different states or one bit per symbol (in other words, 0 = 0, 180 = 1).

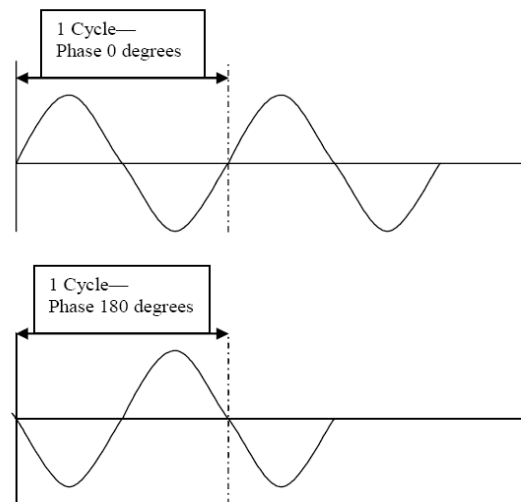

**Figure 2: Phase Shift Keying (PSK)**

In Figure 2, the second wave is shifted by half a period or 180 degrees. The receiver can than recognize this shift indicating either a digital one or zero.

**Quadrature Phase Shifting Keying (QPSK)**

QPSK adds two more phases: 90 and 270 degrees. Now two symbols per bit can be transmitted. Each symbol's phase is compared relative to the previous symbol; so, if there is no phase shift (0 degrees), the bits "00" are represented. If there is a phase shift of 180 degrees, the bits "11" are represented.

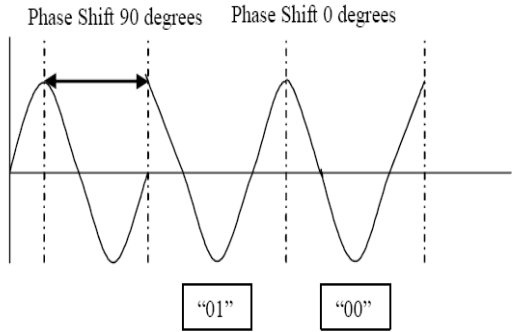

**Figure 3: Quadrature Phase Shift Keying (QPSK)**

| Symbol | Phase Shift |
|--------|-------------|
| 00 | 0 Degrees |
| 01 | 90 Degrees |
| 11 | 180 Degrees |
| 10 | 270 Degrees |

**Table 1: Quadrature Phase Shift Keying (QPSK)**

ASK and PSK can be combined to create QAM where both the phase and amplitude are changed. The receiver then receives this modulated signal, detects the shifts and demodulates the signal back into the original data stream.

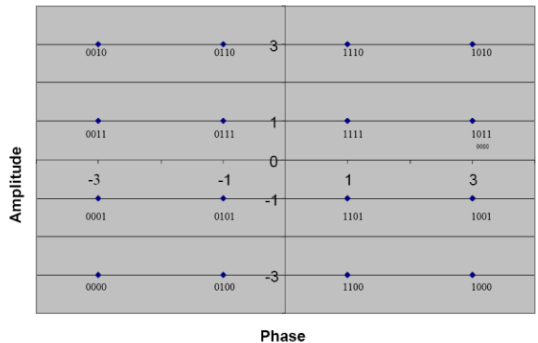

**Figure 4: Quadrature Amplitude Modulation (16-QAM)**

In Figure 4 showing 16-QAM, each symbol can now represent four bits instead of just the two bits per symbol with QPSK. Each point indicates a unique amplitude and phase of the wave (for example, point (1, 1) indicates 90 degrees and amplitude of 1).

**Adaptive Modulation**

Different order modulations allow you to send more bits per symbol and thus achieve higher throughputs or better spectral efficiencies. However, it must also be noted that when using a modulation technique such as 64-QAM, better signal-to-noise ratios (SNRs) are needed to overcome any interference and maintain a certain bit error ratio (BER). The use of adaptive modulation allows a wireless system to choose the highest order modulation depending on the channel conditions.

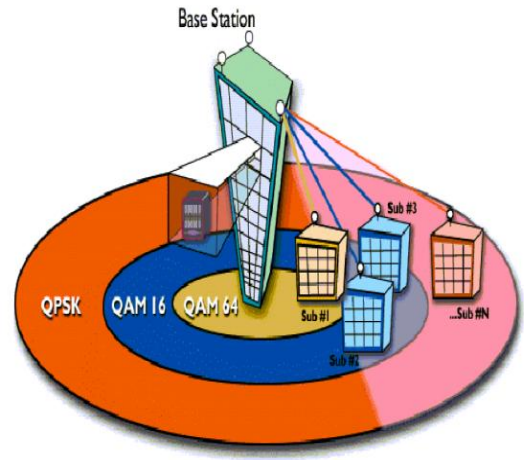

**Figure 5: Adaptive Modulation and Coding**

In Figure 5, you can see a general estimate of the channel conditions needed for different modulation techniques. As you increase your range, you step down to lower modulations (in other words, BPSK), but as you are closer you can utilize higher order modulations like QAM for increased throughput. In addition, adaptive modulation allows the system to overcome fading and other interference.

**Conclusion**

Both QAM and QPSK are modulation techniques used in IEEE 802.11 (Wi-Fi*), IEEE 802.16 (WiMAX*) and 3G (WCDMA/HSDPA) wireless technologies. The modulated signals are then demodulated at the receiver where the original digital message can be recovered. The use of adaptive modulation allows wireless technologies to optimize throughput, yielding higher throughputs while also covering long distances.

**References**

[1] Kundu Sudakshina (2010). Analog and Digital Communications. Pearson Education India. pp. 163–184. ISBN 978-81-317-3187-1.

[2] Ke-Lin Du and M. N. S. Swamy (2010). Wireless Communication Systems: From RF Subsystems to 4G Enabling Technologies. Cambridge University Press. p. 188. ISBN 978-0-521-11403-5.